\documentclass{aastex63}

\usepackage{subfigure}
\shorttitle{ESO~456-SC38}
\shortauthors{Kunder \& Butler}
\begin{document}

\title{Spectroscopic analysis of the bulge Globular Cluster ESO~456-SC38\footnote{Based on observations taken by APOGEE}}

\correspondingauthor{Andrea Kunder, Evan Butler}
\email{akunder@stmartin.edu, embutler2015@gmail.com}

\author[0000-0002-0786-7307]{Andrea M. Kunder}
\affiliation{Saint Martin's University \\
5000 Abbey Way SE \\
Olympia, WA 98503, USA}

\author{Evan Butler}
\affiliation{Saint Martin's University \\
5000 Abbey Way SE \\
Olympia, WA 98503, USA}


\begin{abstract}

It has been suggested that the oldest stellar populations in the Milky Way 
Galaxy are tightly bound and confined to the central regions of the Galaxy.
This is one of the reasons why a handful of globular clusters located in 
the bulge region are thought to be remnants of the primeval 
formation stages of the Milky Way.  The globular cluster, ESO 456-SC38 
(Djorgovski~2) is one such cluster; it has a blue horizontal branch, is 
projected very close to the center of the Galaxy, and 
has an orbit confining it to the bulge/bar region.
The first alpha abundances of seven stars in this heavily reddened cluster 
are presented using APOGEE DR16.  A significant spread in the abundances 
of N, C, Na, and Al indicates the presence of multiple stellar populations 
in this cluster. Using Gaia DR2 proper motions 
and radial velocities from BRAVA-RR, we confirm 
RR Lyrae stars belong to this globular cluster.
\end{abstract}


\section{Introduction} \label{sec:intro}

Globular clusters (GCs) have played a pivotal role in the process of deciphering the 
formation history of the Galaxy \citep[e.g.,][]{brodie06}. These clusters are among the 
oldest objects in the Galaxy, and understanding their formation mechanism sheds light on 
both the formation timescale and conditions of the Milky Way \citep[e.g.,][]{vandenberg13, massari19}.  
Milky Way formation models and an understanding the globular cluster system are linked to 
constraints on the individual stellar components in the 
GCs \citep[e.g.,][]{muratov10, renaud17}.  

Observations of GCs toward the inner Galaxy are notoriously difficult due to high extinction and 
confusion with disk field stars along the line of sight \citep[e.g.,][]{koch17}.  This has contributed to the population of GCs in the Galactic bulge being 
not only understudied but also incomplete \citep[e.g.,][]{bica19}. 
There has been some work focused on obtaining deep and uniform CMDs 
of the bulge GCs \citep[e.g.,][]{minniti17, kerber19, saracino19}, and recent spectroscopic surveys of globular cluster stars 
have also come to the forefront \citep[e.g.,][]{usher17, vasquez18}.  This has filled in 
some of the gaps regarding the distances, chemistry, and kinematics of the bulge clusters. 

It is apparent that the bulge GC population is diverse and harbors the most metal-rich 
GCs in the Galaxy.  
Curiously, it hosts a preponderance of clusters with metallicities of 
$\rm [Fe/H]\sim-$1.0~dex \citep{perezvillegas20}, the majority of which have a blue horizontal 
branch \citep[see e.g.,][]{bica16, cohen18}.  These GCs, as well as their counterpart 
bulge field stars \citep{savino20}, are likely to be some of the oldest objects in the 
Galaxy \citep{barbuy14, barbuy16, barbuy18}.  This is because the older or more metal-poor 
the cluster, the bluer the horizontal branch \citep[see $e.g.,$ Fig.~11 in][]{dias16}.  Therefore, 
metal-rich GCs with a blue HB can only be explained if the cluster has an ancient age. 

This study focuses on the cluster ESO~456-SC38 at ($l$, $b$)=(+2.77, $-$2.50), 
a poorly studied bulge GC with $\rm [Fe/H]\sim-$1.0~dex and a blue horizontal branch.  
Upon its discovery by \citet{holmberg78}, it was classified tentatively as an open cluster.  
Using improved imaging, \citet{djorgovski87} noted that ESO~456-SC38 is in fact a heavily 
obscured globular cluster, and this cluster is therefore also referred to as 
Djorgovski 2 (or Djorg 2).  The first color-magnitude diagram of the cluster was presented 
by \citet{ortolani97} in the optical passbands. They found a distance 
that placed it on the near side of the Galaxy, and the cluster's red giant branch indicated it was 
likely metal-rich with $\rm [Fe/H] \sim -0.5$.  Infrared photometry of this cluster 
presented by \citet{valenti10} yielded similar results, where they found $\rm [Fe/H] = -$0.65~dex and 
$(m-M)_0$ =14.23~mag.  The deepest CMD to date is based on {\it Hubble Space Telescope} 
(HST) photometry \citep{ortolani19}. ESO~456-SC38 was found to have an 
$\rm [Fe/H] = -$1.11~dex with $\rm [\alpha/Fe]=$+0.4, a distance modulus placing it on 
the far side of the bulge with $(m-M)_0$ =14.71$\pm$0.03~mag, and an age of 12.7$\pm$0.7~Gyr.  

Spectroscopic measurements of this cluster have only recently been obtained and have 
been based on a paucity of cluster members.
From four cluster members, \citet{dias16} found ESO 456-SC38 has a 
radial velocity of $-150\pm$28 km s$^{-1}$, $\rm [Fe/H] = -0.79\pm$0.09~dex, and 
$\rm [Mg/Fe] = 0.28 \pm$ 0.10~dex. From three cluster members, \citet{vasquez18} found 
the cluster has a radial velocity of $-$159.9$\pm$0.5 km~s$^{-1}$ and a 
metallicity ranging from $\rm [Fe/H] \sim -$0.97 to $-$1.09~dex, depending on 
the Calcium Triplet calibration used.  Elemental abundances for stars in this cluster 
are largely unknown.  Within the APOGEE~DR16 dataset, we 
were able to isolate seven stars in ESO 456-SC38 with robust metallicities, elemental 
abundances, radial velocities and Gaia DR2 proper motions. This doubles the sample of stars 
with spectroscopic measurements in this cluster and is the first study to shed light 
on a number of elemental abundances for the stars in this cluster.  
Detailed abundance analyses of stars in bulge clusters like 
ESO 456-SC38 are required to determine their origin---for example, whether 
low-latitude, metal-rich clusters formed from 
material that experienced a chemical enrichment history related to bulge stars---
and whether multiple populations exist in these bulge clusters 
\citep[e.g.,][]{schiavon17, johnson18}.

\section{Data and Sample} \label{sec:style}

We use the most recent Apache Point Observatory Galactic Evolution Experiment (APOGEE) 
data release, Sloan Digital Sky Survey-IV DR16 \citep{ahumada20}, to investigate the 
bulge globular cluster ESO 456-SC38. The APOGEE survey \citep{majewski17} collects 
high-resolution (R$\sim$22,500) spectra of stars using near-infrared (NIR) 
wavelengths \citep{wilson19}. The spectra, in general, have 
a signal-to-noise (S/N) that is appropriate for elemental abundance 
determination ($\sim$S/N $>$ 100), and the 
APOGEE Stellar Parameter and Chemical Abundance Pipeline 
\citep[ASPCAP,][]{garciaperez16} provides stellar effective temperatures, 
surface gravities, and metallicities precise to 2\%, 0.1 dex, and 0.05 dex, 
respectively, for most APOGEE stars. This pipeline works especially well for red 
giants, which is the main population targeted by APOGEE.

APOGEE first started collecting data from bulge stars in the Northern Hemisphere, and 
since 2015, APOGEE-2 has begun collecting data from bulge stars in the 
Southern Hemisphere as well. The DR16 contains 473,307 sources with derived atmospheric 
parameters and abundances and is the first data release in which the newer 
APOGEE-2 is made public. 
The APOGEE DR16 catalogue is ideal to search for stars in the cluster 
ESO-456-SC38 due to the dataset's probing of the galactic bulge in the vicinity of 
the cluster. We first isolated all stars in the field containing ESO-456-SC38.
Cluster stars were then selected for membership using the APOGEE
heliocentric radial velocity, the APOGEE
$\rm [Fe/H]$, the projected radial distance from the center of the cluster, and 
Gaia DR2 proper motions.

\begin{figure*}
\centering
\mbox{\subfigure{\includegraphics[height=6.5cm]{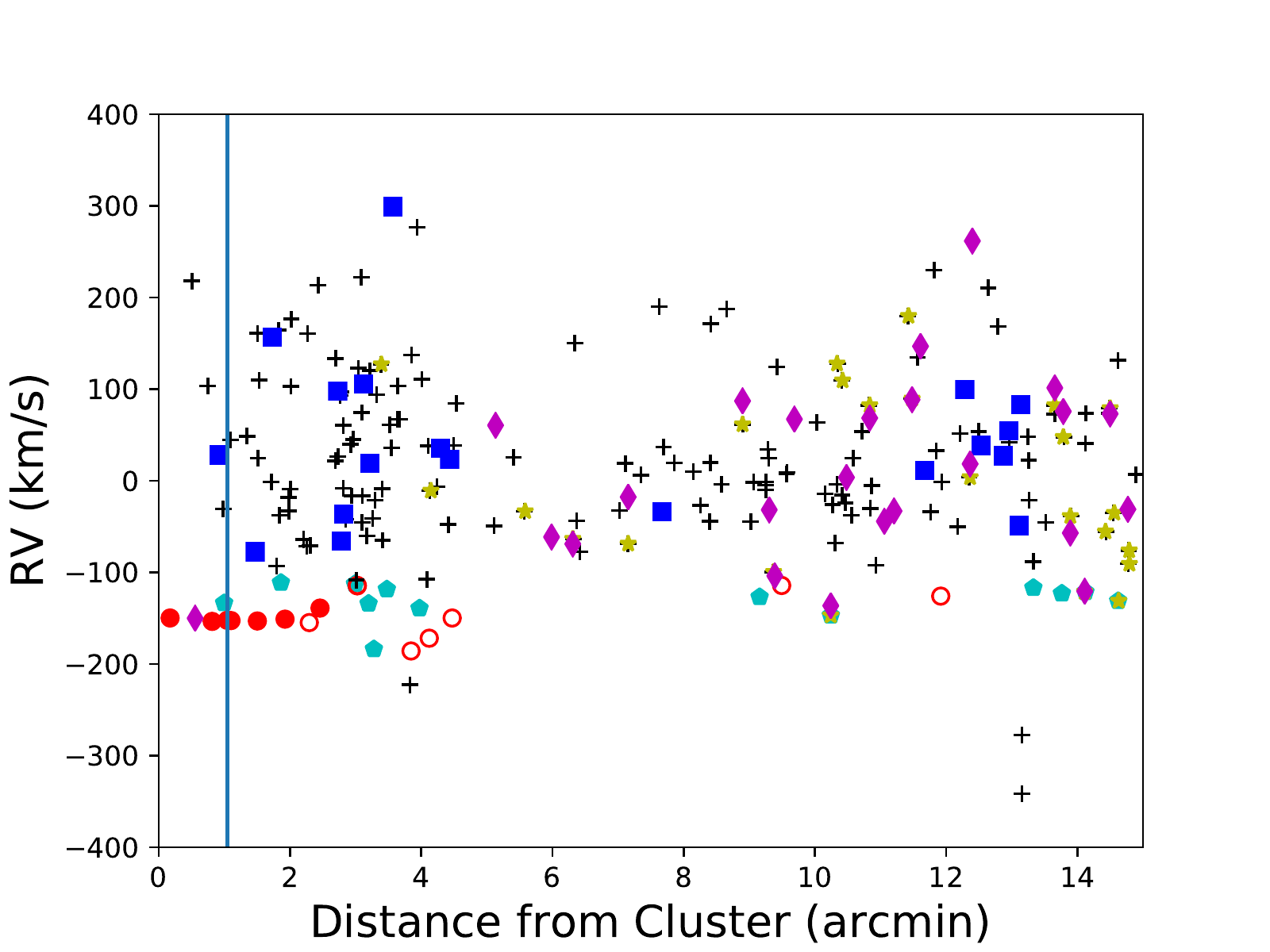}}}
{\subfigure{\includegraphics[height=6.5cm]{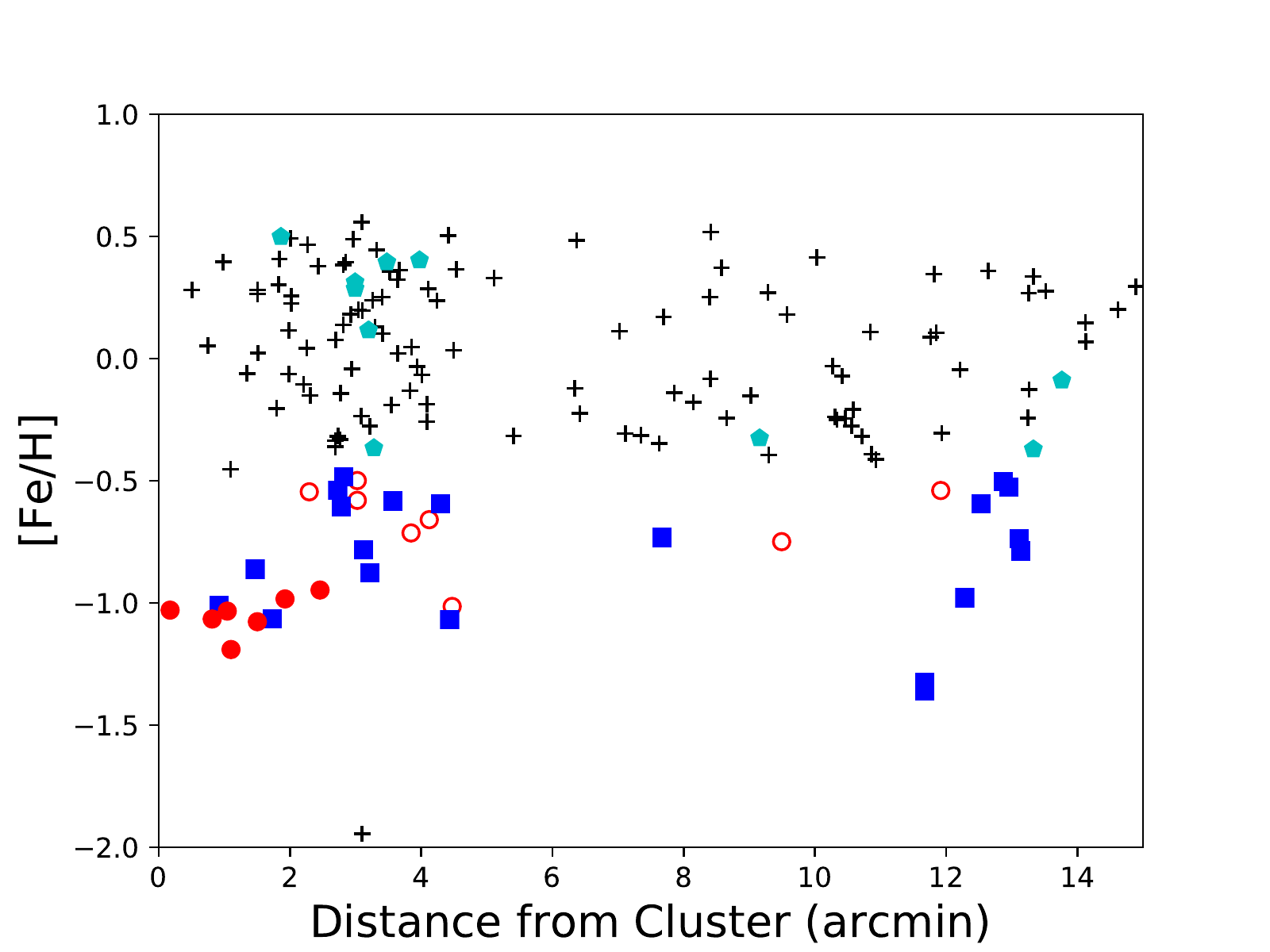}}}
\caption{{\it Left:} Radial distribution of radial velocities in the the field around ESO 456-SC38. The vertical line at 1.05 arcmin illustrates the half-light radius \citet[][2010 edition]{harris96}.  Red circles indicate the member candidates, while hollow red circles indicate possible members. Cyan pentagons show stars close to the GC in radial velocity but not metallicity. Blue squares show stars close to GC in metallicity but not radial velocity. Pink diamonds show RR~Lyrae stars in the BRAVA-RR dataset and gold asterisks show RR~Lyrae stars from the APOGEE dataset. Error bars are the size of the points.
{\it Right:} The $\rm [Fe/H]$ metallicities of the APOGEE stars in the field surrounding ESO 456-SC38. Member candidates are highlighted in red. 
}
\label{spatial}
\end{figure*}

Both the radial velocity and the $\rm [Fe/H]$ metallicity of ESO-456-SC38 
is distinct as compared to the field stars.  The radial velocity 
of ESO-456-SC38 is at least $\sim$50~km~s$^{-1}$ offset from the field, 
and the $\rm [Fe/H]$ metallicity is 
offset by at least 0.5~dex from the field (see Figure~\ref{spatial}). 
Due to the scant number of 
stars in ESO-456-SC38 with spectroscopic measurements, it is not 
immediately obvious what velocity and metallicity range 
is compatible with cluster membership.  
We considered all stars within 15 arcminutes from the cluster
with radial velocities within $\pm$30~km~s$^{-1}$ of the 
mean velocity of $-$150~km~s$^{-1}$ and within 0.5~dex of the mean 
$\rm [Fe/H]$ metallicity of $-$1.0~dex. 

It is apparent that a group of seven stars are offset in both 
velocity and metallicity space, with velocity and metallicity values compatible with 
previous spectroscopic measurements for ESO-456-SC38.  
The velocities of these stars range from $-$139~km~s$^{-1}$ to $-$154~km~s$^{-1}$ 
and the $\rm [Fe/H]$ 
metallicities range from $-$0.95~dex to $-$1.19~dex.  
The proper motion estimate of ESO 456-SC38 reported by \citet{vasiliev19} is ($\overline{\mu_\alpha},\overline{\mu_\delta}$)=(0.515$\pm$0.08~mas~yr$^{-1}$, $-$3.052$\pm$0.08~mas~yr$^{-1}$)
and
all the cluster candidate stars we isolated have proper motions that are within 2$\sigma$ of the 
mean proper motion of the cluster.

The tidal radius of ESO-456-SC38 is reported to be 10.4 arcminutes\citep[from][2010 edition]{harris96}. 
We adopt a limiting cluster radial distance of 15 arcminutes in 
an attempt to identify all APOGEE cluster members belonging to ESO-456-SC38. 
The group of seven stars identified above are within $\sim$3 arcminutes
from the cluster's center.  
We also identify two stars at a radial distance of $\sim$4.5 arcminutes
with radial velocities and metallicities that may indicate they are 
ESO-456-SC38 cluster members.  However, their 
proper motions are greater than 1~mas~yr$^{-1}$ offset from the mean proper motion determined 
for the cluster, 
and so we are uncertain if these stars are bonafide cluster members.  As 
future Gaia data is released, the uncertainties on their stellar proper motions 
will also decrease and we can revisit the range of proper motions expected 
for ESO-456-SC38 cluster membership. 

Figure~\ref{spatial} shows all APOGEE stars 
within 15 arcminutes from the cluster center as a function of their 
radial velocity and $\rm [Fe/H]$ metallicity.  The filled circles indicate 
the ESO-456-SC38 cluster candidates and the possible 
cluster candidates are shown with open circles.  We note that the possible 
cluster candidates, which have $\rm [Fe/H]$ metallicities 
and radial velocities consistent with ESO-456-SC38, do not have proper motions 
that indicate they are moving with the cluster (see Table~\ref{tab:stars}).
%
Table~\ref{tab:stars} gives the APOGEE-ID (column 1), 
APOGEE heliocentric radial velocity (column 2), APOGEE $\rm [Fe/H]$ metallicity (column 3), 
APOGEE $\rm [C/Fe]$, $\rm [N/Fe]$, $\rm [Na/Fe]$, $\rm [Mg/Fe]$, $\rm [Al/Fe]$ (columns 4-8, respectively),
Gaia DR2 proper motion in RA and DEC (column 9 and column 10, respectively), 
distance from the cluster center (column 11), APOGEE effective temperature (column 12), and 
APOGEE surface gravity (column 13) for the ESO-456-SC38 cluster candidates and possible 
candidates.

\begin{figure}
\centering
\mbox{\subfigure{\includegraphics[height=9.0cm, width=20.0cm]{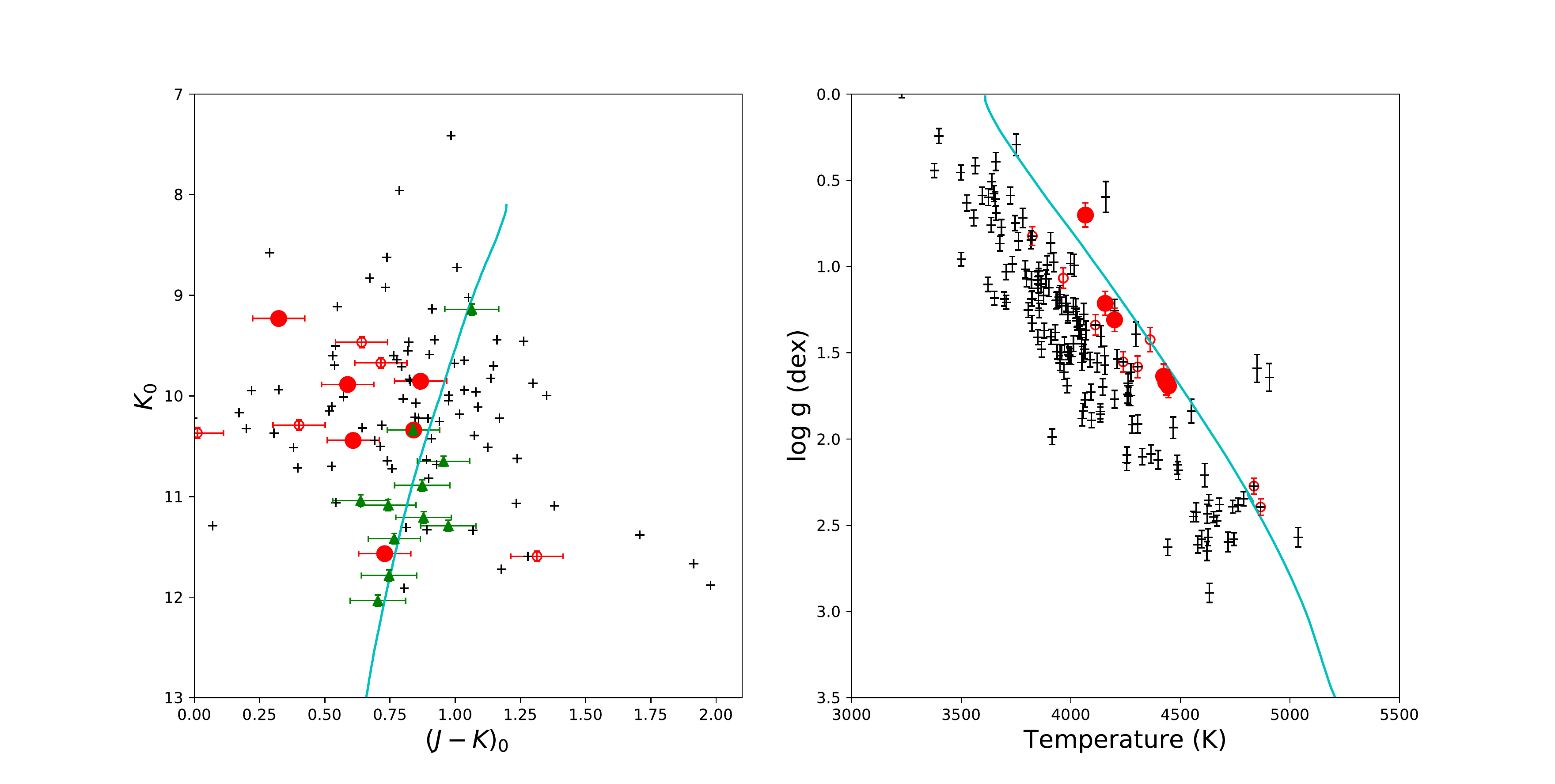}}}
\caption{{\it Left:} Color magnitude diagram from 2MASS photometry for the stars with available APOGEE 
spectra and within $\sim$15 arcmin of ESO 456-SC38.  Member candidates are highlighted in red.  Stars in the Gaia DR2 catalog located within 25" of the cluster are shown as dark green triangles.  A BaSTI isochrone of 12.5 Gyr and alpha-enhanced with $\rm [Fe/H]$ = $-$1.01~dex ($Z$ = 0.004) (cyan line) at a distance modulus of $(m-M)$=14.71 is also over-plotted to show the approximate 
red giant branch of the cluster.
{\it Right:} The log $g$ vs. Teff diagram using the stellar parameters of the same APOGEE stars as 
in left panel.
}
\label{cmd}
\end{figure}

Figure~\ref{cmd} shows the APOGEE stars with a cluster radial distance of 15 arcminutes in 
2MASS $K$ magnitude and $\rm (J-K)$ color 
dereddened using the extinctions from \citet{gonzalez12} and the \citet{nishiyama09} 
extinction law.  From this color-magnitude diagram (CMD), it is apparent that these stars 
populate the cluster's red giant branch, although there is some scatter.  
We speculate the scatter arises from reddening uncertainties, due to e.g., differential 
reddening within the cluster.  This is because the temperatures and surface gravities, which 
are less sensitive to reddening uncertainties, indicate the cluster candidates are on the red giant 
branch of the cluster, as shown in the right panel of Figure~\ref{cmd} (and also see Table~1).

The BaSTI \citep{pietrinferni04, pietrinferni06} alpha-enhanced stellar evolution 
models\footnote{http://basti-iac.oa-abruzzo.inaf.it} were adopted to indicate the approximate 
location of the red giant branch of the cluster.  We used the publicly available BaSTI isochrone 
that best matches the cluster's observed parameters, one with an age of 
12.5 Gyr, a metallicity of $\rm [Fe/H]$ = $-$1.01~dex, and a distance modulus 
of $(m-M)_0$ = 14.71~mag.  This age, metallicity and distance modulus is
consistent with previous parameters determined for the cluster, \citep{ortolani19}.  
Stars from the Gaia DR2 catalogue within 25 arcminutes
of the cluster's center were also added for context.

\begin{figure}
\centering
\mbox{\subfigure{\includegraphics[height=10.0cm, width=17.0cm]{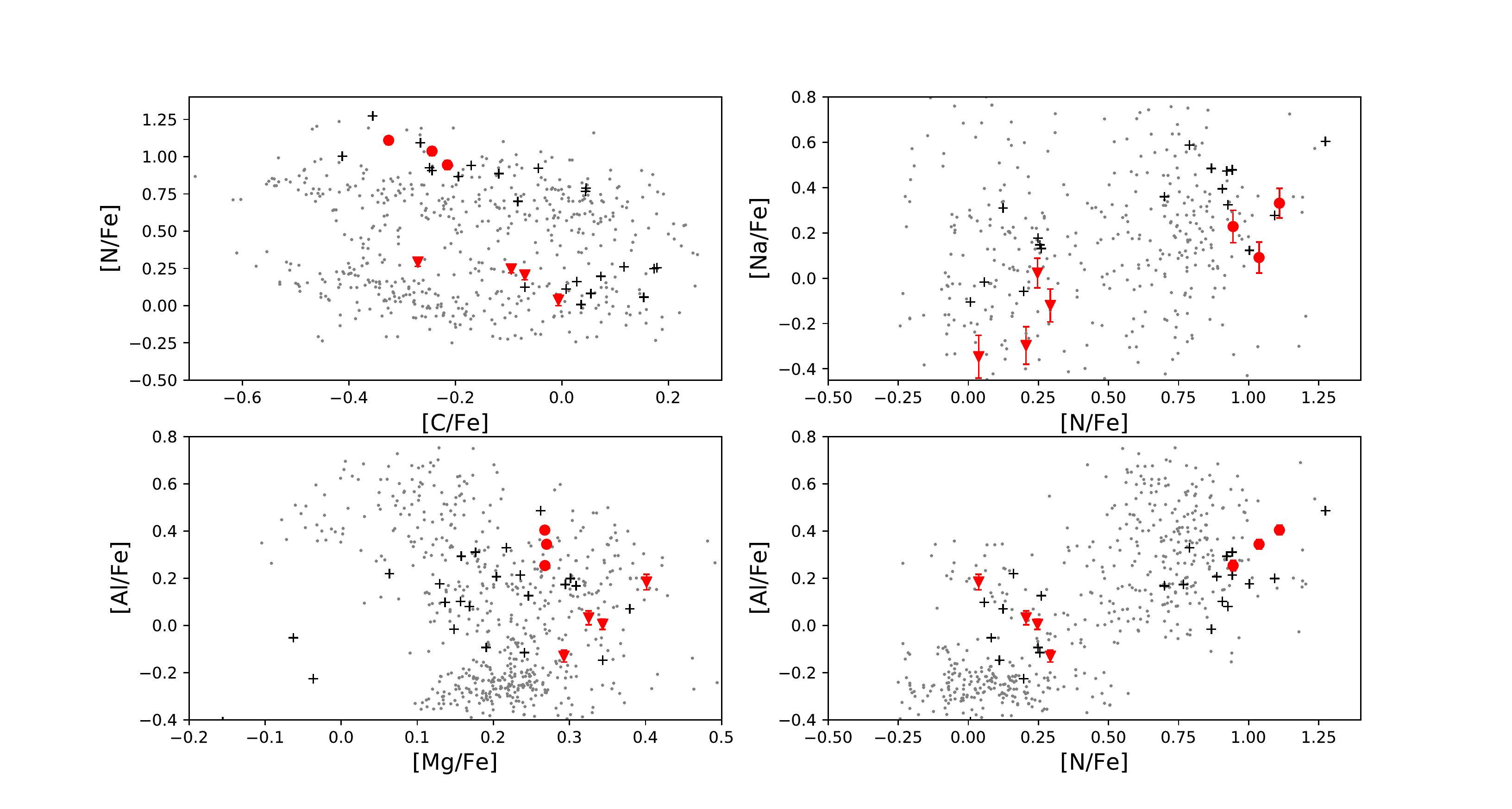}}}
\caption{The abundance ratios $\rm [N/Fe]$, $\rm [Na/Fe]$, $\rm [Al/Fe]$, $\rm [Mg/Fe]$ and 
$\rm [C/Fe]$ for member ESO~456-SC38 stars highlighted in red. Filled circles represent 
N-enhanced stars and inverted triangles show N-normal stars.  Also shown are APOGEE 
abundances from \citet{schiavon17} for stars in the inner Galaxy globular clusters 
NGC~6553, NGC~6528, NGC~6522, Terzan~5 and Palomar~6 in black 
and the stars from \citet{nataf19} in M3, M5, M107, M71 and NGC~6760 in 
grey.}
\label{alpha}
\end{figure}

\section{Results} \label{sec:results}
ESO~456-SC38 is one of the most confined clusters in the innermost Galaxy
with an orbit that keeps it between $R_{GC} \approx$0.15 - 1.67 kpc \citep{perezvillegas20}.
As discussed in the Introduction, this cluster is one of the curious bulge clusters 
belonging to the $\rm [Fe/H] \sim -$1.0 ``peak" with a blue horizontal branch, indicating it is very old 
and formed later than the bar \citep{ortolani19, perezvillegas20}.  Further, 
collisionless and/or hydrodynamic simulations predict that the oldest stars 
reside on tightly bound orbits with small $R_{GC}$ distances \citep[e.g.,][]{tumlinson10, starkenburg17}.  
To study the primeval formation stages of the Milky Way, 
detailed studies of globular clusters such as ESO~456-SC38 are of interest.  

Chemical information in the form of elemental abundance patterns is thought to be 
preserved in stars, and the existence of truly relic stellar groups within the Milky Way 
may be best uncovered using stellar abundances \citep[e.g.,][]{freeman02}.  
For example, open clusters and moving groups exhibit uniform abundance 
patterns \citep[e.g.,][]{bubar10, pancino10} and old globular cluster stars 
now dissolved in the Milky Way have also been uncovered from the 
chemical imprints \citep[e.g.,][]{martell10, schiavon17}.
Unfortunately, very little is known about this cluster from a chemical perspective.  
Spectroscopic observations have been limited to seven cluster members, and the
estimates of $\rm [Fe/H]$ for stars in this cluster range from 
$\rm [Fe/H] = -$0.79 to $-$1.09~dex \citep{dias16, vasquez18}.  The $\rm [Mg/Fe]$ measured 
from low resolution (R$\sim$2000) spectra indicates an $\rm [Mg/Fe] = 0.28 \pm$ 0.10~dex 
for this cluster, but other than this, there have been no reports of elemental abundances of 
this GC.  

Using the APOGEE DR16 database, we have carried out a holistic search for stars 
in ESO~456-SC38, looking simultaneously at radial velocity, $\rm [Fe/H]$, 
and Gaia DR2 proper motion.  This has allowed the identification of cluster stars that were missed previously 
by globular cluster searches in APOGEE-2 \citep[e.g.,][]{meszaros20, horta20}.  
We have doubled the number of ESO~456-SC38 stars with spectroscopic measurements.  
From high-resolution APOGEE spectra (R$\sim$22,500), we find this cluster 
has $\rm [Fe/H] = -1.05 \pm 0.08$~dex where the uncertainty represents the 
scatter about the mean.  This scatter does not seem to indicate the presence of a spread in 
$\rm [Fe/H]$ \citep[e.g.,][]{meszaros20}, although our sample size is small.  

We do see the clear signature of multiple populations within the cluster.  
Figure~\ref{alpha} shows the ESO~456-SC38 stars displayed in various elemental abundance 
planes.  Multiple populations in all globular clusters are evident in the spread of 
C, N, O, and Na in cluster stars \citep[e.g.,][]{kraft94, carretta09}, 
and here a bimodality is the most clearly seen within $\rm [N/Fe]$.  
Three second-generation cluster stars are easily distinguishable in Figure~\ref{alpha}, 
with $\rm [N/Fe]$ abundances of $\sim$1.0~dex and $\rm [Na/Fe]$ abundances of $\sim$0.2~dex. 

Many globular clusters with $\rm [Fe/H]\sim -$1~dex do not show a clear anticorrelation between 
Al and Mg abundances; instead, Mg-Al anticorrelations are typical in clusters with lower 
metallicities where the core temperatures of giant stars are hotter, and so where the Mg-Al
cycle can operate \citep[e.g.,][]{shetrone96, gratton12, meszaros20}. All ESO~456-SC38 stars are 
significantly enhanced in $\rm [Mg/Fe]$ with a $\rm [Mg/Fe]$ scatter of only 0.05~dex. 
The N-enriched stars show $\rm [Mg/Fe]\sim$0.25~dex whereas the 
N-normal stars have $\rm [Mg/Fe] >\sim$0.3~dex. The spread in $\rm [Al/Fe]$ is 0.19~dex, 
with an uncertainty of $\sim$0.05~dex due to the small sample size.  This is compatible 
to the 0.18~dex scatter reported by \citet{meszaros20} for clusters with $-$1.3 $\rm < [Fe/H] < -$1.0~dex.  
The Al-scatter is also could also be consistent with the slightly smaller Al scatter seen for clusters 
with $\rm [Fe/H] > -$1.0~dex \citep[e.g., Fig.~14,][]{meszaros20}, given the small number of stars in 
the cluster and the uncertainties in the APOGEE elemental abundances.  

Because silicon is known to be one of the most reliable $\alpha$-abundance 
measurements in APOGEE \citep{jonsson18}, this element has been used as an indicator 
of the formation of the cluster.  In particular, \citet{horta20} show that at 
metallicities of $\rm [Fe/H] \sim -$1.0~dex, {\it in situ} GC subgroups have 
$\rm [Si/Fe]\sim+$0.25~dex.  In contrast, accreted GCs from accreted subgroups 
such as the Gaia-Enceladus, Helmi streams 
and Sequoia have $\rm [Si/Fe] <$+0.2~dex.  We find that the average $\rm [Si/Fe]$ 
abundances of the ESO~456-SC38 stars is $\rm <[Si/Fe]> =$0.25$\pm$0.06~dex, indicating this 
cluster chemically follows the overall trend of the {\it in situ} GCs.  
However, 
to firmly categorize this cluster, information on its orbit is required \citep[e.g., as in][]{perezvillegas20}.
\\

\begin{figure}
\centering
\mbox{\subfigure{\includegraphics[]{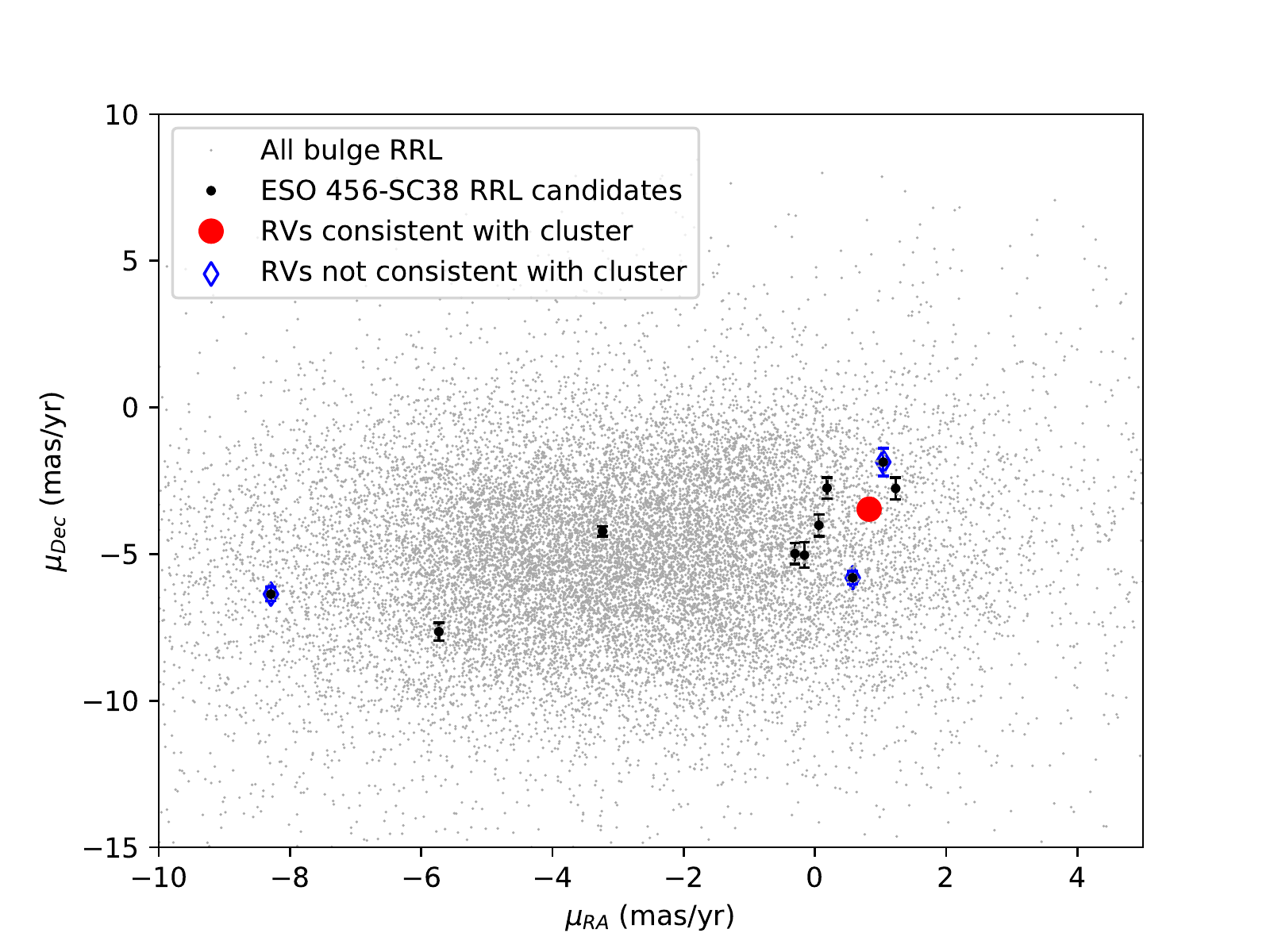}}}
\caption{The Gaia DR2 proper motions of RR Lyrae stars that have been associated with 
ESO~456-SC38 are shown.  Four of these RR Lyrae stars also have radial velocity measurements from 
APOGEE DR16 and/or BRAVA-RR DR2.  A number of RR Lyrae stars associated with the cluster 
are field stars and not cluster members, but a number do have proper motions consistent with 
cluster membership (see Table~\ref{tab:stars}).  }
\label{pm}
\end{figure}

\subsection{RR Lyrae stars}
The horizontal branch of ESO~456-SC38 coupled with its high metallicity indicates this is an 
old cluster of $\sim$12.7$\pm$0.7~Gyr \citep{ortolani19}. Several RR Lyrae stars (RRLs), horizontal branch stars 
residing on the instability strip, have been associated with this cluster.  The extensive 
OGLE-IV \citep{soszynski19} variable star catalog lists 17 RRLs as possible members of 
ESO~456-SC38, and a smaller number of 7 RRLs are tabulated in the 2016 update 
of \citet{clement01}.  Here we attempt to ascertain RRLs that are 
kinematically associated with the cluster.

We crossmatch the OGLE-IV RRL stars with Gaia DR2, finding that 11 of the OGLE-IV stars 
associated with ESO~456-SC38 have a Gaia DR2 counter-part. Figure~\ref{pm} shows proper motions 
of the candidate ESO~456-SC38 RRLs in proper motion space.  Most of these stars do have 
proper motions with values around the proper motion estimate of ESO~456-SC38.  

Using the BRAVA-RR dataset \citep{kunder20}, we verify that 
OGLE-BLG-RRLYR-11190 is a cluster RRL star as it has a 
radial velocity of $-$150 km~s$^{-1}$.  In contrast, OGLE-BLG-RRLYR-11142 has a radial velocity of 
60 km~s$^{-1}$.  This RRL is not moving with the rest of the cluster and is therefore a field star 
instead of a cluster member.  

The APOGEE dataset also includes radial velocities of a number of RRLs.  
Crossmatching APOGEE DR16 with OGLE-IV, we verify that 
OGLE-BLG-RRLYR-11190 is a cluster 
star with a radial velocity of $-$147 km~s$^{-1}$, in good agreement with BRAVA-RR.  
We also find that 
OGLE-BLG-RRLYR-11142 with a radial velocity of 39 km~s$^{-1}$, 
OGLE-BLG-RRLYR-11252 with a radial velocity of 127 km~s$^{-1}$, and 
OGLE-BLG-RRLYR-11372 with a radial velocity of $-$11 km~s$^{-1}$ are field stars instead 
of cluster members.  There are three RRLs with particularly similar proper motions to 
the mean proper motion of 
ESO~456-SC38, but further data, such as radial velocity values, 
would be helpful to confirm membership.  These three RRLs are listed in Table~2 as "Membership 
uncertain" RRLs.

We also find notice there are two RRLs in BRAVA-RR with radial velocities consistent with 
that of ESO~456-SC38, but are at a radial distance of $>$8~arc~min from the 
cluster center (see Figure~\ref{spatial}).  These are OGLE-BLG-RRLYR-10684 and 
OGLE-BLG-RRLYR-11089.  However, their Gaia DR2 proper motions are considerably different that 
what would be expected for cluster membership.

Using the bonafide RRL cluster member, a distance to the cluster can be determined.  
OGLE-BLG-RRLYR-11190 has an average $V$ =  17.848~mag.  The 
absolute magnitude of the star can be found using the Gaia Collaboration et al. 2017 relation:
\begin{equation}
M_V = 0.214~\rm{ [Fe/H]}+0.88.
\end{equation}
Adopting the mean metallicity from the red giant stars presented here 
of $\rm [Fe/H] = -$1.05, this gives $M_V$=0.66~mag.  An identical absolute 
magnitude is found using the theoretical relations from \citet{catelan04}:
\begin{equation}
M_V = -2.288 - 0.882~\rm{log(Z)} + 0.108~log(Z)^2
\end{equation}
where 
\begin{equation}
\rm{log(Z)} = [Fe/H] - 1.765.
\end{equation}
Our uncertainty in $M_V$ is $\sim$0.03~dex, based on our uncertainty in $\rm [Fe/H]$.
Assuming total-to-selective
absorption RV = 3.1, and $\rm E(B-V)$ = 0.81, the distance modulus from 
OGLE-BLG-RRLYR-11190 is $(m-M)_0$ = 14.68$\pm$0.07 (distance of 8.63$\pm$0.28 kpc). 
This distance value is smaller than the distance of 9.12~kpc obtained 
by \citet{ortolani19} using RR Lyrae stars, and instead agrees very well with 
the distance value of 8.75~kpc obtained from CMD fitting by \citet{ortolani19}.
The distance is larger than the 6.3 kpc value in the 
Harris (1996) catalog which 
has recently been adopted for dynamical calculations in \citet{baumgardt19}.

\section{Conclusions \label{conclusions}}
We present 7 red giant stars from APOGEE DR16 that are members of the bulge globular cluster ESO~456-SC38.  
This doubles the sample of stars with spectroscopic measurements for this cluster and 
also adds to the number of stars associated with globular clusters in the APOGEE footprint \citep{horta20, meszaros20}.
From their $\rm [C/Fe]$, $\rm[N/Fe]$, $\rm[Na/Fe]$, $\rm[Mg/Fe]$, and $\rm [Al/Fe]$ abundances, 
we detect the presence of multiple stellar populations in this cluster.  The average $\rm [Si/Fe]$ 
abundances of these stars is $\rm <[Si/Fe]> =$0.25$\pm$0.06~dex, which is typical for {\it in situ} bulge 
clusters \citep[e.g.,][]{horta20} and in agreement with the recent orbit for this cluster from 
\citet{perezvillegas20}.  Using both radial velocities and Gaia DR2 proper motions, 
we show that some RRL that have been associated with the 
cluster are instead field stars.  However, we do confirm the kinematic association
of one RRL and the possible association of three more RRLs,
as expected from its blue horizontal branch.

\acknowledgements
AMK acknowledges support from grant AST-2009836 from the National Science Foundation. 
We thank the anonymous 
referee for suggestions that helped the clarity and quality of the paper.

\rotate
\begin{deluxetable*}{ccccccccccccc}
\tabletypesize{\scriptsize}
\tablenum{1}
\tablecaption{Stellar parameters, elemental abundances, radial velocities and cluster-centric distances for stars in ESO~456-SC38.
\label{tab:stars}}
\tablehead{
\colhead{APOGEE ID} & \colhead{RV} & \colhead{$\rm [Fe/H]$} & \colhead{$\rm [C/Fe]$} & \colhead{$\rm [N/Fe]$} & \colhead{$\rm [Na/Fe]$} & \colhead{$\rm [Mg/Fe]$} & \colhead{$\rm [Al/Fe]$} & \colhead{$\rm \mu_{RA}$} & \colhead{$\rm \mu_{dec}$} & \colhead{$r$} & \colhead{$\rm T_{eff}$} & \colhead{log $g$} 
\\
\colhead{} & \colhead{(km s$^{-1}$)} & \colhead{dex} & \colhead{dex} & \colhead{dex} & \colhead{dex} & \colhead{dex} & \colhead{dex} & \colhead{(mas~yr$^{-1}$)} & \colhead{(mas~yr$^{-1}$)} & \colhead{(arcmin)} & \colhead{(K)} & \colhead{dex} 
\\
}
\decimalcolnumbers
\startdata
\hline
Members\\
\hline
\hline
2M18014557-2750220 & -153.04 $\pm$ 0.01 & -1.08 $\pm$ 0.01 & -0.09 $\pm$ 0.02 & 0.25 $\pm$ 0.03 & 0.02 $\pm$ 0.07 & 0.34 $\pm$ 0.02 & 0.005 $\pm$ 0.02 & 0.93 $\pm$ 0.16 & -3.40 $\pm$ 0.13 & 1.50 & 4157 $\pm$ 77 & 1.21 $\pm$ 0.07\\
2M18014656-2751239 & -138.96 $\pm$ 0.02 & -0.95 $\pm$ 0.02 & -0.006 $\pm$ 0.03 & 0.04 $\pm$ 0.04 & -0.35 $\pm$ 0.09 & 0.40 $\pm$ 0.02 & 0.18 $\pm$ 0.03 & 0.77 $\pm$ 0.20 & -3.45 $\pm$ 0.17 & 2.46 & 4446 $\pm$ 101 & 1.69 $\pm$ 0.07\\
2M18014773-2749465 & -153.48 $\pm$ 0.02 & -1.07 $\pm$ 0.02 & -0.07 $\pm$ 0.02 & 0.21 $\pm$ 0.03 & -0.30 $\pm$ 0.08 & 0.32 $\pm$ 0.02 & 0.03 $\pm$ 0.03 & 0.39 $\pm$ 0.15 & -3.27 $\pm$ 0.12 & 0.81 & 4200 $\pm$ 88 & 1.31 $\pm$ 0.07\\
2M18014786-2749080 & -149.77 $\pm$ 0.01 & -1.03 $\pm$ 0.01 & -0.24 $\pm$ 0.02 & 1.03 $\pm$ 0.03 & 0.09 $\pm$ 0.07 & 0.27 $\pm$ 0.02 & 0.34 $\pm$ 0.02 & 0.52 $\pm$ 0.15 & -2.95 $\pm$ 0.12 & 0.17 & 4433 $\pm$ 83 & 1.67 $\pm$ 0.073\\
2M18015130-2748086 & -152.56 $\pm$ 0.01 & -1.19 $\pm$ 0.02 & -0.27 $\pm$ 0.02 & 0.29 $\pm$ 0.03 & -0.12 $\pm$ 0.07 & 0.29 $\pm$ 0.02 & -0.13 $\pm$ 0.02 & 0.42 $\pm$ 0.13 & -3.06 $\pm$ 0.10 & 1.10 & 4066 $\pm$ 78 & 0.70 $\pm$ 0.07\\
AP18015264-2749084 & -152.27 $\pm$ 0.02 & -1.03 $\pm$ 0.01 & -0.21 $\pm$ 0.02 & 0.94 $\pm$ 0.03 & 0.23 $\pm$ 0.07 & 0.27 $\pm$ 0.02 & 0.25 $\pm$ 0.02 & 0.450 $\pm$ 0.12 & -2.96 $\pm$ 0.10 & 1.04 & 4438 $\pm$ 85 & 1.67 $\pm$ 0.07\\
2M18015592-2749451 & -151.02 $\pm$ 0.01 & -0.98 $\pm$ 0.01 & -0.33 $\pm$ 0.02 & 1.11 $\pm$ 0.03 & 0.33 $\pm$ 0.07 & 0.27 $\pm$ 0.02 & 0.40 $\pm$ 0.02 & 0.63 $\pm$ 0.13 & -3.09 $\pm$ 0.10 & 1.92 & 4423 $\pm$ 82 & 1.64 $\pm$ 0.07\\
\hline
\hline
Membership uncertain \\
\hline
\hline
2M18020645-2750472 & -149.79 $\pm$ 0.02 & -1.01 $\pm$ 0.01 & -0.06 $\pm$ 0.02 & 0.26 $\pm$ 0.03 & -0.10 $\pm$ 0.07 & 0.29 $\pm$  0.02 & 0.02 $\pm$ 0.02 & -2.94 $\pm$ 0.12 & -5.17 $\pm$ 0.10 & 4.47 & 4361 $\pm$ 85 & 1.42 $\pm$ 0.07\\
2M18020649-2748291 & -171.81 $\pm$ 0.01 & -0.66 $\pm$ 0.01 & 0.11 $\pm$ 0.01 & 0.12 $\pm$ 0.02 & 0.09 $\pm$ 0.04 & 0.37 $\pm$ 0.01 & 0.17 $\pm$ 0.02 & 0.33 $\pm$ 0.11& -4.83 $\pm$ 0.09 & 4.12 & 3965 $\pm$ 70 & 1.07 $\pm$ 0.06\\
2M18011802-2742100 & -144.27 $\pm$ 0.02 & -0.75 $\pm$ 0.01 & 0.04 $\pm$ 0.02 & 0.04 $\pm$ 0.03 & -0.17 $\pm$ 0.07 & 0.38 $\pm$ 0.02 & 0.25 $\pm$ 0.03 & -6.26 $\pm$ 0.15 & -6.49 $\pm$ 0.13 & 9.49 & 4305 $\pm$ 88 & 1.58 $\pm$ 0.06\\
2M18012985-2737447 & -125.89 $\pm$ 0.01 & -0.54 $\pm$ 0.01 & 0.17 $\pm$ 0.01 & 0.22 $\pm$ 0.01 & - & 0.34 $\pm$ 0.02 & 0.17 $\pm$ 0.03 & -3.69 $\pm$ 0.16 & -5.45 $\pm$ 0.13 & 11.92 & 3825 $\pm$ 72 & 0.82 $\pm$ 0.05\\
2M18014518-2752453 & -185.77 $\pm$ 0.01 & -0.71 $\pm$ 0.01 & 0.07 $\pm$ 0.01 & 0.08 $\pm$ 0.02 & 0.04 $\pm$ 0.06 & 0.35 $\pm$ 0.02 & 0.10 $\pm$ 0.03 & -6.68 $\pm$ 0.21 & -5.40 $\pm$ 0.16 & 3.85 & 4112 $\pm$ 84 & 1.34 $\pm$ 0.06\\
2M18015255-2751013 & -154.78 $\pm$ 0.01 & -0.55 $\pm$ 0.01 & 0.13 $\pm$ 0.01 & 0.13 $\pm$ 0.02 & 0.04 $\pm$ 0.06 & 0.33 $\pm$ 0.02 & 0.14 $\pm$ 0.03 & -7.73 $\pm$ 0.13 & -4.21 $\pm$ 0.11 & 2.29 & 4239 $\pm$ 89 & 1.55 $\pm$ 0.06\\
2M18015336-2751446 & -144.70 $\pm$ 0.09 & -0.58 $\pm$ 0.02 & 0.06 $\pm$ 0.04 & 0.19 $\pm$ 0.05 & 0.04 $\pm$ 0.11 & 0.67 $\pm$ 0.03 & 0.08 $\pm$ 0.05 & -3.69 $\pm$ 0.23 & 0.69 $\pm$ 0.18 & 3.03 & 4836 $\pm$ 131 & 2.27 $\pm$ 0.05\\
\enddata
\end{deluxetable*}

\begin{deluxetable*}{lllll}
\tabletypesize{\scriptsize}
\tablenum{2}
\tablecaption{RR Lyrae stars in the vicinity of ESO~456-SC38.
\label{tab:rrstars}}
\tablehead{
\colhead{RRL ID} & \colhead{RV} & \colhead{$\rm \mu_{RA}$} & \colhead{$\rm \mu_{dec}$} & \colhead{$r$} \\
\colhead{} & \colhead{(km s$^{-1}$)} & \colhead{(mas~yr$^{-1}$)} & \colhead{(mas~yr$^{-1}$)} & \colhead{(arcmin)} \\
}
\decimalcolnumbers
\startdata
\hline
Member\\
\hline
\hline
OGLE-BLG-RRLYR-11190 & $-$150 $\pm$ 5 & 0.83 $\pm$ 0.31 & -3.47 $\pm$ 0.26 & 0.55 \\
\hline
\hline
Membership uncertain \\
\hline
\hline
OGLE-BLG-RRLYR-11141 & - & 1.23 $\pm$ 0.46 & -2.76 $\pm$ 0.37 & 1.84 \\
OGLE-BLG-RRLYR-11218 & - & 0.18 $\pm$ 0.44 & -2.75 $\pm$ 0.36 & 0.46 \\
OGLE-BLG-RRLYR-11223 & - & 0.06 $\pm$ 0.46 & -4.02 $\pm$ 0.37 & 0.76 \\
\hline
\hline
Non-Member\\
\hline
\hline
OGLE-BLG-RRLYR-11142 & 60 $\pm$ 5 & 0.58 $\pm$ 0.30 & -5.81 $\pm$ 0.22 & 5.13 \\
OGLE-BLG-RRLYR-11252 & 127 $\pm$ 15 & -8.29 $\pm$ 0.29 & -6.37 $\pm$ 0.24 & 3.39 \\
OGLE-BLG-RRLYR-11372 & $-$11 $\pm$ 15 & 1.04 $\pm$ 0.54 & -1.86 $\pm$ 0.48 & 4.15  \\
OGLE-BLG-RRLYR-11049 & - & -3.24 $\pm$ 0.23 & -4.22 $\pm$ 0.17 & 4.70 \\
OGLE-BLG-RRLYR-11060 & - & -5.73 $\pm$ 0.41 & -7.65 $\pm$ 0.30 & 3.81 \\
\enddata
\end{deluxetable*}

{}
\end{document}